\def\bold#1{\setbox0=\hbox{$#1$}%
     \kern-.025em\copy0\kern-\wd0
     \kern.05em\copy0\kern-\wd0
     \kern-.025em\raise.0433em\box0 }
\def\slash#1{\setbox0=\hbox{$#1$}#1\hskip-\wd0\dimen0=5pt\advance
       \dimen0 by-\ht0\advance\dimen0 by\dp0\lower0.5\dimen0\hbox
         to\wd0{\hss\sl/\/\hss}}
\documentstyle[12pt]{article}

\newlength{\dinwidth}
\newlength{\dinmargin}
\newcommand{\resection}[1]{\setcounter{equation}{0}\section{#1}}

\begin{document}
\def\lq{\left [}
\def\rq{\right ]}
\def\LL{{\cal L}}
\def\VV{{\cal V}}
\def\AA{{\cal A}}
\newcommand{\be}{\begin{equation}}
\newcommand{\ee}{\end{equation}}
\newcommand{\bea}{\begin{eqnarray}}
\newcommand{\eea}{\end{eqnarray}}
\newcommand{\nn}{\nonumber}
\newcommand{\dd}{\displaystyle}
\thispagestyle{empty}
\vspace*{4cm}
\begin{center}
  \begin{Large}
  \begin{bf} HADRONIC TRANSITIONS AMONG QUARKONIUM STATES IN A
SOFT-EXCHANGE-APPROXIMATION. CHIRAL BREAKING AND SPIN SYMMETRY BREAKING
PROCESSES.$^*$\\
  \end{bf}
  \end{Large}
  \vspace{5mm}
  \begin{large}
R. Casalbuoni\\
  \end{large}
Dipartimento di Fisica, Univ. di Firenze\\
I.N.F.N., Sezione di Firenze\\
  \vspace{5mm}
  \begin{large}
A. Deandrea, N. Di Bartolomeo and R. Gatto\\
  \end{large}
D\'epartement de Physique Th\'eorique, Univ. de Gen\`eve\\
  \vspace{5mm}
  \begin{large}
F. Feruglio\\
  \end{large}
Dipartimento di Fisica, Univ.
di Padova\\
I.N.F.N., Sezione di Padova\\
  \vspace{5mm}
  \begin{large}
G. Nardulli\\
  \end{large}
Dipartimento di Fisica, Univ.
di Bari\\
I.N.F.N., Sezione di Bari\\
  \vspace{5mm}
\end{center}
  \vspace{2cm}
\begin{center}
UGVA-DPT 12-801\\
hep-ph/9304280\\
December 1992\\
\end{center}
\vspace{1cm}
\noindent
$^*$ Partially supported by the Swiss National Foundation
\newpage
\thispagestyle{empty}
\begin{quotation}
\vspace*{5cm}
\begin{center}
  \begin{Large}
  \begin{bf}
  ABSTRACT
  \end{bf}
  \end{Large}
\end{center}
  \vspace{5mm}
\noindent
Although no asymptotic heavy quark spin symmetry, and even more no flavor
symmetry, are expected for systems such as quarkonium, a numerical discussion
shows that for some processes and in a preasymptotic region which may roughly
include charmonium and bottomonium, the use of the spin-symmetry may be useful
in conjunction with chiral symmetry for light hadrons (soft-exchange-
approximation regime, SEA). We continue our discussion of hadronic
transitions in the SEA-regime by studying in particular chiral breaking
transitions such as $~^3P' \to ~^3P\pi^0$, $~^3P\eta$, level splittings and
transitions which break both chiral and spin symmetry, such as $\psi ' \to
J/\psi \pi^0$, $J/\psi \eta$, and
$~^1P_1 \to J/\psi \pi^0$.
\end{quotation}
\newpage
\setcounter{page}{1}
\resection{Introduction}
The success of heavy quark symmetry \cite{gen}, when applied to systems
containing one heavy quark, does not unfortunately justify its extension to
systems with more than one heavy quark, such as quarkonium, etc. A critical
discussion \cite{noi} of such an issue for quarkonium states has lead us to
recognize that no asymptotic symmetry of the heavy quark type (neither of the
flavor-type nor of spin) is expected to hold for such systems. However at a
numerical level, when limited to a class of processes, and in a preasymptotic
quark mass region, application of a heavy quark formalism, only for the heavy
quark spin symmetry and excluding the flavor symmetry, may be expected to be
of some use. This class of processes excludes those which violate
the Zweig rule and is limited
to a kinematical domain which we have called the SEA
(Soft-Exchange-Approximation) regime, for which it is essential that gluonic
exchanges be predominantly of limited momenta.
\par
The numerical examination of the
possible preasymptotic range for the heavy quark mass suggests that it may be
possible to use the formalism for charmonium and bottomonium, within the
mentioned class of processes, and an effective lagrangian for quarkonia and
light mesons was written down to be used within the SEA regime \cite{noi}.
\par
A number of applications \cite{noi} to transitions among charmonium states, for
not too large momenta of the emitted hadrons, showed the usefulness of the
formalism to derive results which would have otherwise required longer
approximate QCD calculations. The heavy quark spin symmetry alone leads very
simply to general relations for the differential decay rates in hadronic
transitions among quarkonium states, which in the known cases reproduce the
results of a multiple QCD multipole expansion \cite{Yan} for gluonic emission.
Further use of chiral symmetry leads to differential pion decay distributions
valid in the soft regime. As shown in ref. \cite{noi} and \cite{noi1} the
general relations following from heavy quark spin symmetry alone relate the
allowed transitions between two quarkonium multiplets, such as $~^3S_1 \to
{}~^3S_1 + h$ and $~^1S_0 \to ~^1S_0 + h$, $~^3P_2 \to ~^3S_1 + h$, $~^3P_1 \to
{}~^3S_1 + h$, $~^3P_0 \to ~^3S_1 + h$, $~^1P_1 \to ~^1S_0 + h$, all the
transitions $~^3P_2$, $~^3P_1$, $~^1P_1 \to ~^3P_2$, $~^3P_1$, $~^3P_0$,
$~^1P_1 + h$, those of the type $~^3D_3$, $~^3D_2$, $~^3D_1$, $~^1D_2 \to
{}~^3S_1$, $~^1S_0 + h$, independently of the nature of the light final state
$h$. Heavy quark spin symmetry, when supplemented with the lowest order chiral
expansion for the emitted pseudoscalars leads to a general rule allowing only
for even (odd) number of emitted pseudoscalars for transitions between
quarkonium states of orbital angular momenta different by even (odd) units
\cite{noi1}. Such a rule can be violated by higher chiral terms, by chiral
breaking, and by terms breaking the heavy quark spin symmetry. Specialization
to a number of hadronic transitions reproduces
by elementary tensor construction
the known results from the cumbersome multiple expansion in gluon multipoles,
providing for a simple explanation for the vanishing of certain coefficients
which would otherwise be allowed in the chiral expansion. In certain cases,
such as for instance $~^3P_0 \to ~^3P_2 \pi\pi$, $~^3P_1 \to ~^3P_2 \pi\pi$, or
$D-S$ transitions via $2\pi$, the final angular and mass distributions are
uniquely predicted from heavy quark spin and lowest order chiral expansion.
Other processes such as $~^3S_1 \to ~^3S_1 \pi\pi$ will depend on two chiral
parameters, as in the case of $~^3P_0 \to ~^3P_2 \pi\pi$ and $~^3P_1 \to ~^3P_2
\pi\pi$, whereas $~^3P_0 \to ~^3P_1 \pi\pi$ receives no contributions, within
the approximation. We shall not dwell here with the derivation and presentation
of these results for which we refer to \cite{noi1}.
\par
In the present note we
shall concentrate on hadronic transitions among states of quarkonium which
proceed either by breaking of the chiral symmetry, but consistently with heavy
quark spin symmetry, or transitions which break the heavy quark spin symmetry.
For instance the transitions among P-states, $~^3P'_{J'} \to ~^3P_J \pi_0$ or
$\to ~^3P_J \eta$ proceed through chiral breaking but heavy spin conserving
terms, whereas for instance $\psi ' \to J/\psi\pi_0$ or $\to J/\psi\eta$ go
through terms violating both symmetries.
Apart from deriving general relations for the
matrix elements of $\pi^0$ and $\eta^0$ emission in transitions among P-states,
one can estimate the suppression factors entering in these transitions, related
to the current quark masses.
For transitions $\Upsilon (3P_{J'}) \to \Upsilon (1P_J)$ where in the final
state also a $\eta$
could be kinematically allowed, one can estimate the $\pi^0$ versus $\eta$
emission ratio, roughly expected of the order $10^{-2}$ within conventional
assumptions.
\par
 The first test for heavy quark spin breaking is of course in the
structure of levels. The spin breaking in the formalism is expected to go by
insertion of matrices $\sigma_{\mu\nu}$ multiplied by the relevant projectors
at their left and right, and with a depression factor in front of them of the
order
of the inverse of the heavy quark mass. We have tried to reproduce the observed
level patterns, as given by spin-spin, spin-orbit, and tensor splittings, in
terms of $\sigma_{\mu\nu}$ insertions, and found a general consistency for
charmonium and bottomonium, although the lack of flavor symmetry does not allow
for a reliable quantitative comparison between the two systems.
\par
We have studied the transitions $\psi' \to J/\psi \pi_0$ and $\psi' \to
J/\psi \eta$, which go through spin breaking in our formalism, and which are of
interest as the ratio of their partial widths is related to quark masses, apart
 from meson mixing terms.
The transition $~^1P_1 \to J/\psi \pi\pi$ goes through
spin breaking chiral conserving terms, whereas $~^1P_1 \to J/\psi\pi_0$ breaks
both symmetries. In view of a recent upper limit by the E760 collaboration and
of future accurate experiments we have studied both transitions, getting to a
rough estimate of the ratio of their widths, which is in agreement with the
present limits. Increase of experimental accuracies and availability of heavy
meson factories \cite{ki} would make this whole field of experimentation of
renewed interest.
\resection{Discussion of the approximation and formal description}
The usual description of quarkonium states is based on a short-distance regime,
coulomb-like apart radiative corrections, and on a long-distance regime closer
to a string-like description.
\par
A velocity heavy-quark description might make some approximate sense within the
string-like regime, but will certainly fall down in the short distance regime.
For large quark mass the coulombic regime will prevail, and in such a case one
would have from the virial theorem $<T>=-E$, where $T$ is the kinetic energy
and $-E$ the binding energy. Also, from the Feynmann-Hellmann theorem, one
would
have a kinetic energy increasing linearly with the heavy quark mass, implying a
corresponding increase of the relative momentum. The exchange of hard gluons of
large momentum will become dominant. No spin symmetry is then expected to hold.
\par
Even worse for what concerns a possible heavy-quark flavor symmetry. In general
gluon radiation exchanged between static quarks brings about infrared
divergences. In a bound state, potential and kinetic energy play a delicate
balance against each other. The regularization of the infrared divergences then
implies a large breaking of any flavour symmetry because of the explicit
appearance of the heavy quark mass in the kinetic energy.
\par
The conclusion is that, in total contrast to the situation for systems
containing a single heavy quark, no heavy quark spin symmetry emerges
asymptotically for infinite heavy quark mass, and even worse  for a
hypothetical heavy quark flavor symmetry.
\par
For an assessment of the situation in some preasymptotic region one has to
look at the existing quarkonium calculations. We can make use of calculations
of Buchmueller and Tye
\cite{BTye} with a potential behaving like $r^{-1}$ at short distance and like
$r$ at large distance to extract mass behaviours of the type: $<k> \approx
1.0~m_Q^{0.66}$, $<v>\approx~0.5~m_Q^{-0.34}$, $<T>\approx ~0.25~m_Q^{0.32}$
for the residual momentum, the relative velocity and the kinetic energy $T$
within the $Q{\bar Q}$ bound state against the quark mass $m_Q$ expressed in
$GeV$. These formulae are expected to hold at least up to $m_Q\approx ~80~GeV$.
 From calculations by Quigg
and Rosner \cite{QR} for a potential $C~log(r/r_0)$, with $C=3/4~GeV$, we
obtain, by using the virial theorem for this case: $<k>\approx 1.22~m_Q^{0.5}$,
 $<v>\approx 0.65~m_Q^{-0.5}$, $<T>\approx 0.375~GeV$. By applying the
virial theorem and the Feynmann-Hellmann theorem to more recent calculations by
Grant and Rosner \cite{GR} we obtain: $<k>\approx 1.22~m_Q^{0.54}$,
 $<v>\approx 0.61~m_Q^{-0.46}$, $<T>\approx 0.37~m_Q^{0.08}$. The conclusions
seems to be that the kinetic energy and the residual momentum increase with
increasing $m_Q$, while the relative velocity decreases. This seems
empirically true (under all the assumptions for such calculations) in a
preasymptotic region which contains both charmonium and bottomonium. On the
other
hand we know that the asymptotic behaviour for very heavy mass could well imply
a linear growing of the kinetic energy, by naively applying the
Feynmann-Hellmann theorem to a dominant Coulomb force.
\par
For higher waves one finds, using again the analysis of ref.\cite{BTye},
that
for the $c-\bar c$ system the relative velocity increases of about
11-12 \% in going from the s-wave to the d-wave, both for the radial states
$n=1$ and $n=2$. In the case of the bottomonium, the velocity for the $s$, $p$
and $d$ states is almost the same for $n=1$, whereas it increases of about
7\% between the $s$-wave to the $d$-wave for $n=2$. As a consequence we think
that at least up to the $d$-waves, our approach is still consistent.
\par
Once we accept the conjecture that approximate  subasymptotic use of heavy
quark symmetry, limited to the spin symmetry, may be useful in the case of
bottomonium and charmonium, we can easily develop the formalism, following the
notions and the notations of the heavy quark theory as developed for systems
containing one heavy quark. The applications which we had considered in our
previous note \cite{noi} showed no contradictions with existing knowledge and
gave direct and transparent derivations of results which would have required a
lengthy construction of QCD multipole expansion. We summarize here for
completeness and for the notations the description of quarkonium states
\cite{noi}.
\par
A heavy quark-antiquark bound
state,  characterized by  radial number $m$, orbital angular
momentum $l$, spin $s$ and total angular momentum $J$, is
denoted by:
\be
m \,^{2s+1}l_{J}
\label{(2.1)}
\ee
In the limit of no spin-dependent interactions between the two quarks the
singlet $m\,^{1}l_{J}$ and the spin triplet $m\,^{3}l_{J}$  form  a
single multiplet $J(m,l)$. For $l=0$, when the triplet $s=1$
collapses into a single state with total angular momentum $J=1$,
such a multiplet is described by:
\be
J= \frac{(1+\slash v)}{2}[H_{\mu}\gamma^\mu-\eta\gamma_5]
\frac{(1-\slash v)}{2}
\label{(2.4)}
\ee
Here $v^{\mu}$ denotes the four velocity associated to the multiplet $J$;
$H_{\mu}$ and $\eta$ are the spin 1 and spin 0 components respectively; the
radial quantum number has been omitted.
\par
For orbital angular momentum $l\neq 0$ the multiplet $J$ generalizes to
$J^{\mu_1 \ldots \mu_l}$, with a decomposition
\bea
J^{\mu_1 \ldots \mu_l} & = & \frac{(1+\slash v)}{2} \left[
 H_{l+1}^{\mu_1 \ldots \mu_l\alpha} \gamma_{\alpha}
+ \frac{1}{\sqrt{l(l+1)}} \sum_{i=1}^{l} \epsilon^{\mu_i\alpha\beta\gamma}
v_{\alpha} \gamma_{\beta}H_{l\gamma}^{\mu_1 \ldots \mu_{i-1} \mu_{i+1}\ldots
\mu_l} \right. \nn \\
&+& \frac{1}{l}\sqrt{\frac{2l-1}{2l+1}}\sum_{i=1}^{l} (\gamma^{\mu_i}-
v^{\mu_i}) H_{l-1}^{\mu_1 \ldots \mu_{i-1} \mu_{i+1}\ldots \mu_l} \nn \\
 &-&  \frac{2}{l\sqrt{(2l-1)(2l+1)}} \sum_{i<j} (g^{\mu_i\mu_j}-v^{\mu_i}
v^{\mu_j})\gamma_{\alpha} H_{l-1}^{\alpha\mu_1 \ldots \mu_{i-1} \mu_{i+1}\ldots
\mu_{j-1}\mu_{j+1}\ldots \mu_l} \nn \\
&+&  K_l^{\mu_1\ldots\mu_l} \gamma_5 \left ] \frac{(1-\slash v)}{2} \right.
\label{(2.5)}
\eea
Here $K_l^{\mu_1\ldots\mu_l}$ represents the spin singlet
$\,^1 l_J$, and
the spin triplet $\,^3 l_J$ is represented by
$H_{l+1}^{\mu_1 \ldots \mu_{l+1}}$ for $J=l+1$,
$H_{l}^{\mu_1 \ldots \mu_l}$ for $J=l$, and
$H_{l-1}^{\mu_1 \ldots \mu_{l-1}}$ for $J=l-1$. All these tensors are
completely symmetric, traceless and satisfy transversality conditions
\bea
& & v_{\mu_1}K_l^{\mu_1\ldots\mu_l} = 0  \nn \\
& & v_{\mu_1}H_{l+1,l,l-1}^{\mu_1\ldots\mu_{l+1,l,l-1}}  = 0
\label{(2.6)}
\eea
Moreover, to avoid orbital momenta
other than $l$, we require that $J^{\mu_1 \ldots \mu_l}$ itself is
completely symmetric, traceless and orthogonal to the velocity.
This allows to identify the states in \ref{(2.5)} with the physical states.
The normalisation for $J^{\mu_1 \ldots \mu_l}$ has been chosen so that:
\bea
< J^{\mu_1 \ldots \mu_l}{\bar J}_{\mu_1 \ldots \mu_l} > =
& 2 & \left( H_{l+1}^{\mu_1 \ldots \mu_{l+1}}
H^{\dagger l+1}_{\mu_1 \ldots \mu_{l+1}}
-H_{l}^{\mu_1 \ldots \mu_{l}}H^{\dagger l}_{\mu_1 \ldots \mu_{l}}\right. \nn \\
& + & \left. H_{l-1}^{\mu_1 \ldots \mu_{l-1}}H^{\dagger l-1}_{\mu_1 \ldots
\mu_{l-1}}-K_{l}^{\mu_1 \ldots \mu_{l}}K^{\dagger l}_{\mu_1 \ldots \mu_{l}}
\right)
\label{(2.8)}
\eea
where ${\bar J}=\gamma^0 J^\dagger \gamma^0$ and $< \ldots >$ means the trace
over the Dirac matrices.
The following applications will concern only $s$ and $p$  states, given
respectively by \ref{(2.4)} and
\bea
J^\mu = \left. \frac{1+\slash v}{2} \right [
   & H_2^{\mu\alpha}\gamma_{\alpha} &  +
      \frac{1}{\sqrt{2}}\epsilon^{\mu\alpha\beta\gamma}v_\alpha\gamma_\beta
H_{1\gamma}  \nn \\
&+& \left. \frac{1}{\sqrt{3}} (\gamma^\mu-v^\mu)H_0 +K_1^\mu \gamma_5 \right]
\frac{(1-\slash v)}{2}
\label{(2.9)}
\eea
Under a Lorentz transformation $\Lambda$ we have:
\be
J^{\mu_1 \ldots \mu_{l}} \rightarrow \Lambda^{\mu_1}_{~\nu_1} \ldots
\Lambda^{\mu_l}_{~\nu_l} D(\Lambda) J^{\nu_1 \ldots \nu_l} D(\Lambda )^{-1}
\label{lorentz}
\ee
where $D(\Lambda)$  is the usual spinor representation of $\Lambda$.
\par
Parity and charge conjugation have the following action:
\bea
J^{\mu_1 \ldots \mu_{l}} & \stackrel{P}{\rightarrow} &
\gamma^0 J_{\mu_1 \ldots \mu_{l}} \gamma^0
\label{2.12}
\eea
\bea
J^{\mu_1 \ldots \mu_{l}} & \stackrel{C}{\rightarrow} &
(-1)^{l+1} C J^{\mu_1 \ldots \mu_{l}T} C
\label{2.13}
\eea
where $C=i\gamma^2\gamma^0$ is the charge conjugation matrix.
\par
Under heavy quark spin transformation one has
\be
J^{\mu_1 \ldots \mu_l} \rightarrow S J^{\mu_1 \ldots \mu_l} S'^\dagger
\label{2.19}
\ee
with $S,\,S' \in SU(2)$ and $[S,\slash v]= [S',\slash v]=0$.
As long as one can neglect spin
dependent effects, one will require invariance of the allowed interaction terms
under the transformation \ref{2.19}.
\resection{Chiral breaking hadronic transitions}
In this note we restrict ourselves  to hadronic transitions with
emission of light pseudoscalar mesons. Such a light sector, in the limit of
vanishing quark masses, has a spontaneously broken
$SU(3) \times SU(3)$ chiral symmetry. The light pseudoscalar octet is
described in terms of pseudo-Goldstone bosons, assembled in the matrix
\be
\Sigma=\exp{ \frac{2iM}{f_\pi}}
\label{3.1}
\ee
where $f_\pi$ is the pion decay constant, $f_\pi \simeq 132 \, MeV$,
and
\be
{M}=
\left (\begin{array}{ccc}
\sqrt{\frac{1}{2}}\pi^0+\sqrt{\frac{1}{6}}\eta & \pi^+ & K^+\nn\\
\pi^- & -\sqrt{\frac{1}{2}}\pi^0+\sqrt{\frac{1}{6}}\eta & K^0\\
K^- & {\bar K}^0 &-\sqrt{\frac{2}{3}}\eta
\end{array}\right )
\label{3.2}
\ee
Frequently occurring quantities are the 1-forms $ {\cal A}_\mu$ and
${\cal V}_\mu$, given by:
\bea
{\cal A}_\mu & = & \frac{1}{2} \left( \xi^\dagger \partial_\mu \xi -
\xi\partial_\mu \xi^\dagger \right) \nn \\
{\cal V}_\mu & = & \frac{1}{2} \left( \xi^\dagger \partial_\mu \xi +
\xi\partial_\mu \xi^\dagger \right)
\label{3.3}
\eea
with $\xi^2=\Sigma$.
\par
 Under the chiral symmetry the fields $\xi$ and $\Sigma$
transform as follows:
\bea
\xi & \to  & g_L \xi U^\dagger=U \xi g_R^\dagger \nn \\
\Sigma & \to & g_L \Sigma g_R^\dagger
\label{3.4}
\eea
where $g_L,\, g_R$ are global $SU(3)$ transformations and $U$ is a function
of $x$, of the fields, and of $g_L$, $g_R$.
The forms ${\cal A}_\mu$ and ${\cal V}_\mu$ transform as:
\bea
{\cal A}_\mu &\to & U {\cal A}_\mu U^\dagger\nn\\
{\cal V}_\mu &\to & U {\cal V}_\mu U^\dagger+U\partial_\mu U^\dagger
\label{3.5}
\eea
Under parity we have:
\bea
\Sigma & \stackrel{P}{\rightarrow} & \Sigma^\dagger \nn \\
{\cal A}_\mu &\stackrel{P}{\rightarrow} & -{\cal A}^\mu \nn \\
{\cal V}_\mu &\stackrel{P}{\rightarrow} & {\cal V}^\mu
\label{3.6}
\eea
Under charge conjugation:
\bea
\Sigma & \stackrel{C}{\rightarrow} & \Sigma^T \nn \\
{\cal A}_\mu &\stackrel{C}{\rightarrow} & {\cal A}^T_\mu \nn \\
{\cal V}_\mu &\stackrel{C}{\rightarrow} & -{\cal V}^T_\mu
\label{3.7}
\eea
In this section we will discuss possible chiral breaking but spin
conserving terms, which are important for transitions forbidden in the $SU(3)
\times SU(3)$ symmetry limit. Examples of such kind of transitions are
\be
^3P_{J'} \to \;^3P_J \pi_0, \; \;^3P_J \eta
\label{6.1}
\ee
The transitions
\be
\psi' \to J/\psi \pi_0, \, J/\psi \eta
\label{6.3}
\ee
need terms which in addition violate the spin symmetry and will be discussed
in the next section.
\par
We first discuss the masses and mixings of the octet and singlet $\eta'$
pseudoscalar light meson states. The term which give mass to the pseudoscalar
octet, massless in the chiral limit, is
\be
{\cal L}_{m}=\frac{\mu f_{\pi}^2}{4} <M (\Sigma + \Sigma^{\dagger})>
\label{6.4}
\ee
Here $M$ is the current mass matrix:
\be
{M}=
\left (\begin{array}{ccc}
m_u & 0 & 0 \nn\\
0 & m_d & 0  \\
0 & 0 & m_s
\end{array}\right )
\label{6.5}
\ee
and $\mu$ is a scale parameter with dimensions of a mass. The lagrangian
\ref{6.4} gives in addition a mixing $\pi_0 - \eta$: the physical states
${\tilde \pi}_0, \; {\tilde \eta}$ turn out to be:
\bea
{\tilde \pi}_0 & = & \pi_0 +\epsilon \eta \nn \\
{\tilde \eta} & = & \eta- \epsilon \pi_0
\label{6.6}
\eea
where the mixing angle $\epsilon$ is
\be
\epsilon=\frac{(m_d -m_u) \sqrt{3}}{4 (m_s-\dd\frac{m_u+m_d}{2})}
\label{6.7}
\ee
The $\eta'$, which is a chiral singlet, mixes with
$\pi_0, \; \eta$.
Such a mixing can be described by the term
\be
{\cal L}_{\eta\eta'}=\frac{i f_\pi}{4} \lambda <M (\Sigma - \Sigma^{\dagger})>
\eta'
\label{6.8}
\ee
where $\lambda$ is a parameter with dimension of a mass. At first order in the
mixing angles  the physical states are:
\bea
{\tilde \pi}_0 & = & \pi_0 +\epsilon \eta +\epsilon' \eta' \nn \\
{\tilde \eta} & =  & \eta -\epsilon \pi_0 +\theta \eta' \nn \\
{\tilde \eta}' & = & \eta' -\theta \eta -\epsilon' \pi_0
\label{6.9}
\eea
where
\bea
\epsilon' & = & \frac{\lambda (m_d -m_u)}{\sqrt{2} (m^2_{\eta'}- m^2_{\pi_0})}
\nn \\
\theta  & = & \sqrt{\frac{2}{3}}~~~\frac{\lambda \dd\left(
m_s-\frac{m_u+m_d}{2}
\right)}{m^2_{\eta'}- m^2_{\eta}}
\label{6.10}
\eea
and $\epsilon$ as given in \ref{6.7}.
\par
We will consider chiral violating, spin-conserving hadronic transitions between
charmonium states at first order in the chiral breaking mass matrix
$M$. We are thus lead to consider the quantities:
\bea
& &<M (\Sigma +\Sigma^{\dagger})>   \nn \\
& &<M (\Sigma - \Sigma^{\dagger})>
\label{6.11}
\eea
The first one is even under parity, the second odd, and both
have $C=+1$.
\par
The only term spin-conserving and of leading order in the current quark masses
contributing to the transition \ref{6.1} is
\be
<J_\mu {\bar J}_\nu > v_\rho \epsilon^{\mu\nu\rho\sigma} \partial_\sigma
\left[ \alpha \frac{i f_\pi}{4} <M (\Sigma - \Sigma^{\dagger})> + \beta
f_\pi\eta'
\right]
\label{6.12}
\ee
where $\alpha$ and $\beta$ are coupling constants of dimensions
$(mass)^{-2}$.
The direct coupling to $\eta'$ contributes through the mixing \ref{6.9}.
The spin symmetry of the heavy sector gives relations among the modulus square
matrix elements of the transitions between the two $p$-wave states. In
particular we find  that
\be
|{\cal M}|^2 (^3P_0 \to ^3P_0\pi)
=|{\cal M}|^2 (^3P_2 \to ^3P_0\pi)=0
\label{6.13}
\ee
and that all non-vanishing matrix elements can be expressed in terms of
$^3P_0 \to ^3P_1\pi$:
\bea
|{\cal M}|^2 (^3P_1 \to ^3P_1\pi)& =&\frac{1}{4}
|{\cal M}|^2 (^3P_0 \to ^3P_1\pi)   \nn \\
|{\cal M}|^2 (^3P_1 \to ^3P_2\pi)& = &\frac{5}{12}
|{\cal M}|^2 (^3P_0 \to ^3P_1\pi)   \nn \\
|{\cal M}|^2 (^3P_2 \to ^3P_2\pi)& = &\frac{3}{4}
|{\cal M}|^2 (^3P_0 \to ^3P_1\pi)   \nn \\
|{\cal M}|^2 (^1P_1 \to ^1P_1\pi)& =&
|{\cal M}|^2 (^3P_0 \to ^3P_1\pi)
\label{6.14}
\eea
where $\pi$ stays for $\pi_0$ or $\eta$. The relations \ref{6.14} can be
generalized for any spin conserving transition between $l=1$ multiplets,
leading to the same results of a  QCD double
multipole expansion \cite{Yan}
\par
The width for the emission of a $\pi_0$ follows from \ref{6.12}:
\be
\Gamma (^3 P_0 \to ^3P_1 \pi_0)= \frac{3}{8 \pi}|\vec{p}_\pi
 |^3\left(m_d-m_u\right)^2\left[\alpha+\frac{2}{3}\beta\frac{\lambda
f_\pi}{\left(m_{\eta'}^2-m_{\pi_0}^2\right)}\right]^2
\label{6.15}
\ee
where $\vec{p}_\pi$ is the momentum of the emitted pion in the rest frame of
the decaying particle.
The width is suppressed  approximately
by a factor $(m_u -m_d)^2/\Lambda^2$ where $\Lambda=\Lambda_{QCD}$.
For most of the transitions between P-states there is not enough phase space
for the emission of an $\eta$. However a $\eta$ could be observed for
$\Upsilon(3 P_J)$ going to $\Upsilon( 1 P_J)$.
For such transitions the ratio of the partial widths turns out to be:
\be
\frac{\Gamma (^3 P_J \to ^3 P_{J'}' \pi_0 )}{\Gamma (^3 P_J \to ^3 P_{J'}'
\eta)}=\frac{27}{16} \left| \frac{\vec{p}_\pi}{\vec{p}_\eta} \right|^3
 \left[ \frac{m_d -m_u}{
m_s-\frac{m_u+m_d}{2}} \right]^2
\left[\frac{1+\dd\frac{2}{3}\frac{\beta}{\alpha}\frac{\lambda f_\pi}
{\left(m_{\eta'}^2-m_{\pi_0}^2\right)}}
{1+\dd\frac{\beta}{\alpha}\frac{\lambda f_\pi}
{\left(m_{\eta'}^2-m_{\eta}^2\right)}}
\right]^2
\label{6.16}
\ee
By assuming a small direct coupling of the $\eta '$ ($\beta\ll\alpha$),
or by neglecting
the mixing $\pi_0 -\eta'$ and $\eta -\eta'$ (small $\lambda$),
we can estimate the previous
ratio.
Taking $m_d -m_u = 5 \,MeV$, $m_s= 150\, MeV$, and the mass of $\Upsilon(3P_J)$
equal to 10.53 $GeV$, as predicted in potential models \cite{BTye},
  one has for the ratio \ref{6.16} the value:
\be
R=1.3 \times 10^{-2}
\label{6.18}
\ee
\def\qpiu{Q_v^{(+)}}
\def\qmeno{Q_v^{(-)}}
\def\propiu{\frac {1+\slash v}{2}}
\def\promeno{\frac {1-\slash v}{2}}
\def\opiu{O_1^{(+)}}
\def\omeno{O_1^{(-)}}
\section{Spin breaking}
For heavy mesons there are only two types of operators that can break
spin symmetry. The simple reason is that on the quark (antiquark) indices
of the quarkonium wave function act
projection operators $(1+\slash v)/2$ and $(1-\slash v)/2$ which
reduce the original $4\times 4$-dimensional space to a $2\times 2$-dimensional
one. Obviously, in the rest frame, the most general
spin symmetry breaking term is of the form $\vec a\cdot\vec\sigma$,
where $\vec\sigma$ are the Pauli matrices.
In an arbitrary frame one observes that any $\Gamma$-matrix
sandwiched between  two projectors $(1+\slash v)/2$, or $(1-\slash v)/2$,
can be reexpressed in terms of $\sigma_{\mu\nu}$ sandwiched between the
same projectors:
\bea
\frac{1+\slash v}{2} 1\frac{1+\slash v}{2}&=&\frac{1+\slash v}{2}\\
\frac{1+\slash v}{2}\gamma_5\frac{1+\slash v}{2}&=&0\\
\frac{1+\slash v}{2}\gamma_\mu\frac{1+\slash v}{2}&=&v_\mu\frac{1+\slash
v}{2}\\
\frac{1+\slash v}{2}\gamma_\mu\gamma_5\frac{1+\slash v}{2} &=&
\frac{1}{2}\epsilon_{\mu\nu\alpha\beta}v^\nu\frac{1+\slash v}{2}
\sigma^{\alpha\beta}\frac{1+\slash v}{2}\\
\frac{1+\slash v}{2}\gamma_5\sigma_{\mu\nu}\frac{1+\slash v}{2} &=&
-\frac{i}{2}\epsilon_{\mu\nu\alpha\beta}\frac{1+\slash v}{2}
\sigma^{\alpha\beta}\frac{1+\slash v}{2}
\eea
and analogous relations with $(1+\slash v)/2\to(1-\slash v)/2$.
We use here $\epsilon_{0123}=+1$. Let us define
\be
\sigma_{\mu\nu}^{(\pm)}=\frac{1 \pm \slash v}{2}\sigma_{\mu\nu}
\frac{1 \pm \slash v}{2}
\ee
In the rest frame, $\sigma_{\mu\nu}^{(\pm)}$ reduce to Pauli matrices.
 From the previous identities it follows that the most general spin symmetry
breaking terms in the quarkonium space are of the form
$G_1^{\mu\nu}\sigma_{\mu\nu}^{(+)}$, or $G_2^{\mu\nu}\sigma_{\mu\nu}^{(-)}$,
with $G_i^{\mu\nu}$ two arbitrary antisymmetric tensors. There is another
convenient way to express this result by means of the Pauli-Lubanski
four-vector
\be
\Sigma_\mu=\frac{1}{4}\epsilon_{\mu\nu\alpha\beta}v^\nu\sigma^{\alpha\beta}=
\frac{i}{2}\gamma_5\sigma_{\mu\nu}v^\nu
\ee
In fact, due to the following identity, $\sigma_{\mu\nu}^{(+)}$
can be evaluated in terms of $\Sigma_\mu$
\be
\sigma_{\mu\nu}^{(+)}=2\epsilon_{\mu\nu\alpha\beta}\frac{1+\slash v}{2}
\Sigma^\alpha v^\beta\frac{1+\slash v}{2}
\ee
$\Sigma_\mu$ is orthogonal to $v_\mu$, and in the rest frame
we have $\Sigma^\mu=(0,\vec\sigma/2)$.
\par
We have shown that the most general operators which break spin
symmetry are  $\sigma_{\mu\nu}^{(\pm)}$. We can try to get
some more insight at the problem by looking at the underlying QCD theory.
Following \cite{Grins}, we analyze the QCD equations of motion of
a heavy quark
\be
(i\slash D-M)\psi(x)=0
\ee
where $D=\partial+ig_s G$, and $G$ the gluon field. Introducing the velocity
dependent fields
\be
\psi(x)=e^{-iMv\cdot x}Q_v
\ee
we get for the projections
\be
Q_v^{(\pm)} = \frac{1\pm\slash v}{2} Q_v
\ee
the following equations of motion:
\be
iv\cdot D \qpiu = -\propiu i\slash D\qmeno,~~~~~~~
\left(1+\frac{iv\cdot D}{2M}\right)\qmeno =
\frac{1}{2 M}\promeno i\slash D\qpiu
\ee
We can solve formally the equation for $\qmeno$
\be
\qmeno=\frac{1}{2M}\frac{1}{\displaystyle{1+\frac{iv\cdot D}{2M}}}
\frac{1-\slash v}{2}i\slash D\qpiu
\ee
and substitute the result
into the first equation, obtaining
\be
iv\cdot D\qpiu=-\frac{1}{2M}\propiu i\slash D\promeno\frac{1}{\displaystyle{
1+\frac{iv\cdot D}
{2M}}}
i\slash D\propiu\qpiu
\ee
The price for eliminating $\qmeno$ is a non-local equation for $\qpiu$.
However, the usefulness of the previous equation is in the expansion in $1/M$.
By using the identity
\be
\propiu\gamma_\mu\promeno\gamma_\nu\propiu=\propiu\left(g_{\mu\nu}-v_\mu v_\nu-
i\sigma_{\mu\nu}\right)\propiu
\ee
we get our final result
\be
iv\cdot D\qpiu=\frac{1}{2M}\propiu\left[g_{\mu\nu}-v_\mu v_\nu-i
\sigma_{\mu\nu}\right]D^\mu\frac{1}{\displaystyle{1+\frac{iv\cdot D}{2M}}}
D^\nu\qpiu
\ee
In particular, at the first order in $1/M$ we have
\be
iv\cdot D\qpiu=\frac{1}{2M}\propiu\left[D^2-(v\cdot D)^2+
\frac{g_s}{2}G_{\mu\nu}\sigma^{\mu\nu}\right]\qpiu
\ee
where
\be
G_{\mu\nu}=\frac{1}{ig_s}\left[D_\mu,D_\nu\right]
\ee
In the rest frame, this equation is nothing but the Pauli-Schr\"odinger
equation.
We could also use the relation between $\qmeno$ and $\qpiu$ in the
equation of motion for the gluons, to obtain
that the only spin symmetry
breaking term is proportional to $\sigma_{\mu\nu}^{(+)}$, with the
further information that the coefficient of this operator starts with $1/M$.
 From this argument we expect that any insertion of the operator
$\sigma_{\mu\nu}^{(+)}$ gives a suppression factor $1/M$.
Analogous conclusions can be reached  for $\sigma_{\mu\nu}^{(-)}$
by considering the heavy anti-quark field.
\par
The first example of spin breaking within the formalism will concern
the fine structure of $J^{\mu_1 \ldots \mu_{\ell}}$ levels in
a few interesting cases. The general expression for the
fine structure in terms of spin and angular
momentum consists of a linear combination of
\be
a=\bf {S_1} \cdot \bf {S_2}
\ee
\be
b=\bf {L} \cdot \bf {S}
\ee
\be
c=-\frac {1}{(2\ell -1) (2\ell +3)} [ 12 ({\bf {L}} \cdot {\bf {S}})^2
+ 6 {\bf {L}} \cdot {\bf {S}} -4 {\bf {S}}^2 {\bf {L}}^2]
\ee
where $\bf {S_1}$ and $\bf {S_2}$ are the quark spins,
$\bf {S}$ is the total spin of the system, and $\bf {L}$ its orbital
angular momentum. The first term gives the hyperfine splitting, the second the
spin-orbit splitting and the third comes from the tensor term.
The
corresponding matrix elements for $S$, $P$, $D$ states of quarkonium are
given in table I.
Within our formalism, for the S-wave, the hyperfine
splittings arise from the following term:
\be
A(S) = <\sigma^{\mu \nu} J \sigma_{\mu \nu} {\overline J}>
\ee
The values of table I, column a, are reproduced with an appropriate
numerical coefficient
in front of this term and recalling our normalization for the S-wave of
eq.\ref{(2.8)}.
In the case of the P-wave, the spin-spin, spin-orbit, and tensor terms, are
given respectively by:
\be
A(P) = <J^{\mu} \sigma_{\nu \rho} {\overline J}_{\mu} \sigma^{\nu \rho}>
\ee
\be
B(P) = i <J^{\mu} \sigma_{\mu \nu} {\overline J}^{\nu}>
- i <{\overline J}^{\nu} \sigma_{\mu \nu} J^{\mu}>
\ee
\be
C(P) = <J^{\mu} \sigma_{\mu \nu} {\overline J}_{\rho} \sigma^{\rho \nu}> +
<J^{\mu} \sigma^{\rho \nu} {\overline J}_{\rho} \sigma_{\mu \nu}>
\ee
where the last term is in effect a combination of the usual tensor and
spin-spin terms.  The
analogous terms for D-waves are:
\be
A(D) = <J^{\mu \nu} \sigma_{\rho \lambda} {\overline J}_{\mu \nu}
\sigma^{\rho \lambda}>
\ee
for the spin-spin term,
\be
B(D) = i <J^{\mu \nu} \sigma_{\mu \rho} {\overline J}_{\nu}^{\rho}>
- i <{\overline J}_{\nu}^{\rho} \sigma_{\mu \rho} J^{\mu \nu}>
\ee
for the spin-orbit term, and
\be
C(D) = <J^{\mu \nu} \sigma_{\mu \rho} {\overline J}_{\lambda \nu} \sigma^{\rho
\lambda}> + <J^{\mu \nu} \sigma^{\rho \lambda} {\overline J}_{\lambda \nu}
\sigma_{\mu \rho}>
\ee
\be
C'(D) = <J^{\mu \nu} \sigma^{\mu \rho} {\overline J}_{\rho \lambda}
\sigma_{\nu \lambda}>
\ee
which are both  combinations of the usual tensor and spin-spin terms. It is
easy to build a linear combination of the two which gives the tensor
splittings of table I.
\par
 From the mass values of table II one can extract the physical
splittings and perform a fit to arrive at
the numerical coefficients in front of our
lagrangian terms. As the coefficients have the dimensions of  mass, we choose
them to be in $MeV$ and obtain for the S-wave spin-spin splitting $-7.3 A(S)$
and $-2.5 A(S)$ for $c{\overline c}$ and $b{\overline b}$ respectively, in
front of
an unperturbed mass levels of $1534 <J{\overline J}>$ and $4730
<J{\overline J}>$ for the two cases.
\par
For P-waves the corresponding coefficients are $2 A(P)$ and
$0.8 A(P)$, to be compared with the unsplitted common mass term of $1762.6
<J^{\mu}{\overline J}_{\mu}>$ and $4950.1\break\noindent
 <J^{\mu}{\overline J}_{\mu}>$
for charmonium and bottomonium respectively.
For the splittings within the triplets, the mass spectra are reproduced by the
following combinations:
\be
8.75 B(P) - 2 [A(P) - 1.5 C(P)]
\ee
in the case of $c{\overline c}$ and
\be
3.5 B(P) - 0.6 [A(P) - 1.5 C(P)]
\ee
in the case of $b{\overline b}$.
We see that, in agreement with our previous considerations, the spin-spin and
the tensor terms are depressed with respect to the spin-orbit coupling, which
contains a single $\sigma_{\mu\nu}$ insertion. Also these terms are more
depressed for bottomonium than for charmonium.
In the previous computation we have assumed
that possible mixing terms among different waves are negligible. It is however
possible, within the formalism, to include such mixing terms in the
lagrangian. For
example in the case of a $S-D$ wave mixing, such a term is given by
\be
<J \sigma_{\mu \rho} {\overline J}^{\mu \nu} \sigma_{\nu}^{\rho}> +
<{\overline J} \sigma_{\mu \rho} J^{\mu \nu} \sigma_{\nu}^{\rho}>
\ee
mixing $~^3S_1$ and $~^3D_1$ states.
\resection{Spin breaking hadronic transitions}
We apply now our formalism to the transitions $\psi' \to J/\psi \pi_0$ and
$\psi' \to J/\psi \eta$. Of particular interest is the ratio
\be
R=\frac {\Gamma (\psi' \to J/\psi \pi_0)}{\Gamma(\psi' \to J/\psi \eta )}
\label{7.1}
\ee
which provides for a measure of the light-quark mass ratio
\be
r=\frac {m_d - m_u}{m_s -\dd\frac{m_u+m_d}{2}}
\label{7.2}
\ee
Using partial conservation of axial-vector current Ioffe and Shifman
\cite{Ioffe} give the prediction
\be
R=\frac{27}{16} \left[\frac{\vec{p_{\pi}}}{\vec{p_{\eta}}}\right]^3 r^2
\label{7.3}
\ee
The calculation of $R$ is straightforward with the heavy quark formalism. Eq.
\ref{7.3} will be recovered when neglecting the mixings $\pi_0 - \eta$ and
$\eta -\eta'$ (or a possible direct coupling of $\eta'$).
\par
The most general spin breaking lagrangian
 for the processes $\psi' \to J/\psi \pi_0 , \eta$  is
\bea
{\cal L}=&i\epsilon_{\mu\nu\rho\lambda} \left[ <J'\sigma^{\mu\nu}\bar{J}> -
<\bar{J}\sigma^{\mu\nu}J'>\right] v^{\rho} \times \nn\\
&\partial^{\lambda} \left[\dd\frac{i A}{4} <M(\Sigma -\Sigma^{\dagger})>
+B\eta ' \right] +h.c.
\label{7.4}
\eea
The couplings $A$ and $B$ have dimension $(mass)^{-1}$; the $B$ term
contributes to the ratio \ref{7.1}
 via the mixing $\pi_0 -\eta '$ and $\eta -\eta '$,
in the same way as the $\beta$ coupling in \ref{6.12}. There are no terms
with the insertion of two $\sigma$; the two P and C conserving  candidates
\bea
&\epsilon_{\mu\nu\rho\lambda} \left[
<J'\sigma^{\mu\tau}\bar{J}\sigma_{\tau}^{~\nu}> +
<\bar{J}\sigma^{\mu\tau}J'\sigma_{\tau}^{~\nu}>\right] v^{\rho}
\partial^{\lambda} <M(\Sigma -\Sigma^{\dagger})>;\nn\\
&\epsilon_{\mu\nu\rho\lambda} \left[
<J'\sigma^{\mu\nu}\bar{J}\sigma^{\rho\lambda}> +
<\bar{J}\sigma^{\mu\nu}J'\sigma^{\rho\lambda}>\right]
<M(\Sigma -\Sigma^{\dagger})>
\label{7.5}
\eea
are both vanishing. Using the lagrangian \ref{7.4} and taking into account the
mixings \ref{6.9} we can calculate the ratio \ref{7.1},
 which is quite similar to the
ratio \ref{6.16}
\be
R=\frac{27}{16} \left[\frac{\vec{p_{\pi}}}{\vec{p_{\eta}}}\right]^3
\left[ \frac {m_d - m_u}{m_s -1/2 (m_u+m_d)}\right]^2 \left[\frac{
1+\dd\frac{2 B}{3 A}\frac{\lambda f_{\pi}}{m^2_{\eta '}-m^2_{\pi0}}}
{1+\dd\frac{B}{A}\frac{\lambda f_{\pi}}{m^2_{\eta '}-m^2_{\eta}}}\right]^2
\label{7.6}
\ee
If we neglect the mixings $\pi_0 -\eta '$ and $\eta -\eta '$ ($\lambda =0$) or
the direct coupling of $\eta '$ (B=0) \ref{7.6} reduces to \ref{7.3}.
\par
Eq. \ref{7.6} can receive corrections from electromagnetic
contributions to the transition $\psi ' \to J/\psi \pi_0$. It has been
shown that such corrections are suppressed \cite{Don}, \cite{Malt}.
A second type
of corrections is associated with higher order terms in the light-quark mass
expansion (the lagrangian \ref{7.4} is the first order of such  expansion); a
discussion can be found in ref. \cite{DW}.
\par
We consider now two hadronic decay modes for the recently discovered
\cite{E760} $^1P_1$ state of charmonium. These processes are:
\bea
^1P_1 & \to & J/\psi \pi\pi \label{7.7} \\
^1P_1 & \to & J/\psi \pi_0
\label{7.8}
\eea
The first one is, at the leading order, spin breaking but chiral conserving,
while the second one breaks both symmetries. Therefore one could naively expect
an enhancement of \ref{7.7} respect to \ref{7.8}. Voloshin \cite{volo}
suggested that the isospin violating transition \ref{7.8}
 could be an order of magnitude stronger
than the two pion transition \ref{7.7}.
Kuang Tuan and Yan \cite{kty} prediction is quite different,
but the E760 Collaboration \cite{E760} has set the upper limit
\be
\frac {\Gamma (^1P_1\to~J/\psi\pi\pi)}{\Gamma (^1P_1\to~J/\psi \pi_0)}
\leq 0.18
\label{7.9}
\ee
This result is consistent only with the prediction of Voloshin.
We now give an estimate of the partial widths for these decays in our approach.
\par
For the decay $~^1P_1\to~J/\psi~\pi\pi$ we can write down in general
the following terms:
\bea
& &a \left[<J_\mu\sigma^{\mu\nu}\bar{J}>+<\bar{J}\sigma^{\mu\nu} J_\mu>\right]
<{\cal A}_\nu(v\cdot {\cal A})>\\
& & i~b  \left[<J^\mu\sigma_{\mu\rho}\bar{J}\sigma^{\rho\nu}>-<\bar{J}
\sigma_{\mu\rho} J^\mu \sigma^{\rho\nu}>\right]
<{\cal A}_\nu(v\cdot {\cal A})>\nn
\eea
with $a$ and $b$ arbitrary coefficients with dimension $mass^{-1}$.
The contributions of the "heavy" factors of
these operators to the matrix element of the process under study are
of the same form so that one obtains
\be
4 (a+b) \epsilon^{\alpha\beta\nu\delta} H_\alpha K_\beta v_\delta
\ee
where $H$ is the field describing the $J/\psi$ resonance and $K$ the $^1P_1$.
\par
For the decay $~^1P_1\to~J/\psi \pi_0$ we can list three independent terms
in the lagrangian
\bea
& & i~c~  \epsilon_{\mu\nu\rho\sigma}
\left[<J^\mu\sigma^{\nu\rho}\bar{J}>+<\bar{J}\sigma^{\nu\rho} J^\mu>\right]
v^\sigma<M\Sigma-M\Sigma^\dagger>;\label{cde}\\
 & & d~ \epsilon_{\mu\nu\rho\sigma}
\left[<J^\mu\sigma_{\tau}\,^{\rho}\bar{J}\sigma^{\nu\tau}>-<J^\mu
\sigma^{\nu\tau}\bar{J}\sigma_{\tau}\,^{\rho}>\right]
v^\sigma<M\Sigma-M\Sigma^\dagger>;\nn\\
 & & e~\epsilon_{\mu\nu\rho\sigma}
\left[<\sigma^{\mu\tau}J_{\tau}\sigma^{\nu\rho}\bar{J}>-<\sigma^{\nu\rho}
J_{\tau}\sigma^{\mu\tau}\bar{J}>\right]
v^\sigma<M\Sigma-M\Sigma^\dagger> \nn
\eea
with $c,d,e$ arbitrary dimensionless coefficients.
 The contribution of the heavy part
to the matrix element of the process $~^1P_1\to~J/\psi~\pi_0$ sums up to
\be
8 (c+d+2e) H \cdot K
\ee
For the ratio of the partial widths for $~^1P_1\to~J/\psi~\pi_0$ and
for $~^1P_1\to~J/\psi~\pi\pi$ we find:
\be
\frac {\Gamma (^1P_1\to~J/\psi \pi\pi)}{\Gamma (^1P_1\to~J/\psi \pi_0)}=
1.7\times 10^{-2} \left( \frac{a+b}{c+d+2e} \right)^2 ~GeV^2
\label{ratio}
\ee
Due to our ignorance of the coefficient ratio in \ref{ratio} we cannot give
an exact prediction. However one can try an estimate.
If we use our previous argument about $\sigma_{\mu\nu}^{(\pm)}$
insertions, we expect
$a$ to be of order $1/M_c$ and $b$ of order $1/M_c^2$. On the other hand,
the coefficients in \ref{cde} are expected to be proportional to
$\Lambda_\chi$  ($\Lambda_\chi \approx 1~GeV$), and furthermore we
expect $c$ of order $1/M_c$  and $d$, $e$ of order $1/M_c^2$. Therefore,
except for possible cancellations,
it seems reasonable to assume:
\be
\frac{a+b}{c+d+2e} \approx \frac{1}{\Lambda_\chi}
\ee
leading to a rough estimate
\be
\frac {\Gamma (^1P_1\to~J/\psi \pi\pi)}{\Gamma (^1P_1\to~J/\psi \pi_0)} \approx
2 \times 10^{-2}
\ee
This would provide for a possible explanation of the relative suppression
 of the  two pion
channel.
Finally we notice that in our approach the decays $~^1P_1\to~J/\psi~\pi\pi$
and $~^1P_1\to~J/\psi~\pi_0$ can be related to other processes:
$^3P_0 \to ~^1S_0 \pi_0$ and $~^3P_1 \to ~^1S_0 \pi\pi$. An analogous estimate
in the case of bottomonium leads to:
\be
\frac {\Gamma (^1P_1\to~\Upsilon \pi\pi)}{\Gamma (^1P_1\to~\Upsilon \pi_0)}
\approx 2.6 \times 10^{-2}
\ee
\resection{Conclusions}
There is no theoretical basis for heavy-quark symmetry as asymptotic symmetry
of bound quarkonium in the limit of infinite quark mass. Nevertheless, in a
preasymptotic region, expectedly including charmonium and bottomonium,
numerical discussion, plus a number of successful applications, show the
practical usefulness of adopting the heavy-quark formalism to describe a
certain class of processes, which do not violate the Zweig rule, and
furthermore only by limiting to (broken) heavy-quark spin symmetry, that is
excluding heavy-quark flavor symmetry. The processes are characterized by
predominantly soft gluon-exchanges (SEA regime), both for the essential of the
bound state description and for the occurring dynamical gluon exchanges. The
usefulness of the description appears in particular in conjunction with use of
the chiral expansion for the light pseudoscalars, allowing for the construction
of an effective chiral lagrangian for charmonium states and light
pseudoscalars, which has been successfully applied to study of hadronic
transitions \cite{noi}, \cite{noi1}.
\par
In this note we have explored, within such an approach, transitions which
proceed either by breaking the chiral symmetry, or by breaking the heavy-quark
spin symmetry, or both symmetries. For processes such as $~^3P'_{J'} \to ~^3P_J
\pi^0$ or $~^3P'_{J'} \to ~^3P_J \eta^0$ one can relate among them $\pi^0$ and
$\eta^0$ emission and estimate the suppression factors in terms of the current
quark masses. A manifestation of the heavy quark spin breaking is in the
observed level splittings, reproduced in this approach through spurion-type
spin-breaking insertions, and corresponding to the standard spin-spin,
spin-orbit, and tensor splittings. We have also calculated the partial widths
of the heavy-quark spin-breaking transitions $\psi' \to J/\psi \pi^0$ and
$\psi' \to J/\psi \eta^0$, whose ratio is directly related to quark masses, and
the p-state to s-state transitions $~^1P_1 \to J/\psi \pi \pi$ and
$~^1P_1 \to J/\psi \pi^0$, both spin-breaking, the first one chiral-conserving,
the second one chiral-violating. Recent and forthcoming improvements in the
experimental limits and expected future availability of heavy meson factories
will make comparison with data more precise and theoretically informative.
\newpage
\begin{center}
\begin{Large}
\begin{bf}
Table Captions
\end{bf}
\end{Large}
\end{center}
\vspace{5mm}
\begin{description}
\item {\bf Table I}: Matrix elements for spin-spin (a),
spin-orbit (b), and tensor term (c) in S, P, D states of quarkonium.
\item {\bf Table II}: Masses (in MeV) of S and P states of charmonium and
bottomonium. All values are experimental, except $~^1P_1$ and $~^1S_0$ states
for bottomonium \cite{potenziale}. Data are from \cite{PDG}, except for
$~^1P_1$ state of charmonium \cite{E760}.
\end{description}

\newpage
\begin{table}
\begin{center}
\begin{tabular}{l c c c }
& {\bf Table I} & & \\ & & & \\
\hline \hline
${\rm State}$ & a & b & c  \\ \hline
$~^1S_0$ & -3/4 & 0 & 0 \\ \hline
$~^3S_1$ & 1/4 & 0 & 0 \\ \hline
$~^1P_1$ & -3/4 & 0 & 0 \\ \hline
$~^3P_0$ & 1/4 & -2 & -4 \\ \hline
$~^3P_1$ & 1/4 & -1 & 2 \\ \hline
$~^3P_2$ & 1/4 & 1 & -2/5 \\ \hline
$~^1D_2$ & -3/4 & 0 & 0 \\ \hline
$~^3D_1$ & 1/4 & -3 & -2 \\ \hline
$~^3D_2$ & 1/4 & -1 & 2 \\ \hline
$~^3D_3$ & 1/4 & 2 & -4/7 \\ \hline
\hline
\end{tabular}
\end{center}
\end{table}
\vskip 1truecm
\begin{table}
\begin{center}
\begin{tabular}{l c c }
& {\bf Table II} & \\ & & \\
\hline \hline
${\rm State}$ & $c{\overline{c}}$ & $b{\overline{b}}$ \\ \hline
$~^1S_0$ & 2980 & 9420 \\ \hline
$~^3S_1$ & 3097 & 9460 \\ \hline
$~^1P_1$ & 3526 & 9901 \\ \hline
$~^3P_0$ & 3415 & 9860 \\ \hline
$~^3P_1$ & 3510 & 9892 \\ \hline
$~^3P_2$ & 3556 & 9913 \\ \hline
\hline
\end{tabular}
\end{center}
\end{table}
\end{document}